\documentclass[conference]{IEEEtran}
\IEEEoverridecommandlockouts
\usepackage{cite}
\usepackage{amsfonts}
\usepackage{algorithmic}
\usepackage{textcomp}
\usepackage{balance}
\usepackage{color}
\usepackage{fancybox}
\usepackage{graphicx}
\usepackage{cellspace}
\usepackage{enumitem}
\usepackage{comment}
\usepackage{diagbox}
\usepackage{url}
\usepackage{xspace}
\usepackage{amssymb}
\usepackage{multirow}
\usepackage{cite}
\usepackage[caption=false]{subfig}
\usepackage{epsfig}
\usepackage{cellspace}
\usepackage[english]{babel}
\usepackage[utf8x]{inputenc}
\usepackage[T1]{fontenc}
\usepackage{amsmath}
\usepackage{tikz}
\usepackage{xcolor}
\usepackage{graphicx}
\usepackage{caption}
\usepackage{subcaption}
\usepackage{float} %
\graphicspath{{images/}}
\usepackage{adjustbox} 
\usepackage{tcolorbox}
\usepackage{listings}
\usepackage{pifont}
\usepackage{graphicx}
\usepackage{float} 
\usepackage{booktabs} 
\usepackage{titlesec}
\usepackage{titletoc}
\usepackage{float}
\usepackage{microtype} 
\usepackage{amsthm}
\usepackage{pifont}
\usepackage{makecell}

\usepackage{siunitx}

\def\BibTeX{{\rm B\kern-.05em{\sc i\kern-.025em b}\kern-.08em
    T\kern-.1667em\lower.7ex\hbox{E}\kern-.125emX}}
\begin{document}

\title{CARGO: A Framework for Confidence-Aware Routing of Large Language Models}


\author{
\IEEEauthorblockN{Amine Barrak\IEEEauthorrefmark{1}, 
Yosr Fourati\IEEEauthorrefmark{1}\IEEEauthorrefmark{3}, 
Michael Olchawa\IEEEauthorrefmark{1}, 
Emna Ksontini\IEEEauthorrefmark{2}, 
Khalil Zoghlami\IEEEauthorrefmark{1}}
\IEEEauthorblockA{\IEEEauthorrefmark{1}\textit{Department of Computer Science and Engineering, Oakland University}, Rochester, MI, USA}
\IEEEauthorblockA{\IEEEauthorrefmark{2}\textit{University of North Carolina Wilmington}, Wilmington, NC, USA}
\IEEEauthorblockA{\IEEEauthorrefmark{3}\textit{Mediterranean Institute of Technology (MEDTECH)}, Tunis, Tunisia \\
Email: aminebarrak@oakland.edu}
}

\maketitle

\begin{abstract}
As large language models (LLMs) proliferate in scale, specialization, and latency profiles, the challenge of routing user prompts to the most appropriate model has become increasingly critical for balancing performance and cost. We introduce \textbf{CARGO} (\emph{Category-Aware Routing with Gap-based Optimisation}), a lightweight, confidence-aware framework for dynamic LLM selection. CARGO employs a single embedding-based regressor trained on LLM-judged pairwise comparisons to predict model performance, with an optional binary classifier invoked when predictions are uncertain. This two-stage design enables precise, cost-aware routing without the need for human-annotated supervision. To capture domain-specific behavior, CARGO also supports category-specific regressors trained across five task groups: mathematics, coding, reasoning, summarization, and creative writing. Evaluated on four competitive LLMs (GPT-4o, Claude 3.5 Sonnet, DeepSeek V3, and Perplexity Sonar), CARGO achieves a top-1 routing accuracy of \textbf{76.4\%} and win rates ranging from \textbf{72\%} to \textbf{89\%} against individual experts.

These results demonstrate that confidence-guided, lightweight routing can achieve expert-level performance with minimal overhead, offering a practical solution for real-world, multi-model LLM deployments.
\end{abstract}

\vspace{0.2cm}

\section{Introduction}
\label{sec:introduction}

Large language models (LLMs) have advanced significantly in recent years, driven by improvements in model architectures, training methodologies, and large-scale data availability \cite{zhao2023survey}. These developments have expanded their capabilities beyond traditional natural language processing (NLP) tasks, allowing complex reasoning, code generation, knowledge retrieval, and multimodal understanding \cite{multimodal}. Modern LLMs incorporate billions of parameters, self-attention mechanisms, and reinforcement learning to enhance response quality based on human feedback \cite{shao2024survey}. 

However, LLMs are optimized for different objectives, some excel in efficiency and factual accuracy, while others prioritize creativity and depth of reasoning \cite{joshi2024strategic}. For instance, models tailored for coding tasks emphasize precise syntax generation and debugging assistance \cite{chen2024survey}, whereas those optimized for creative writing focus on coherence and stylistic variation \cite{gomez2023confederacy}. These differences arise from variations in training data, alignment techniques, and external knowledge integration. 

As LLMs are deployed across diverse applications, users submit queries that span domains such as mathematics, coding, creative writing, and business analysis. Each domain imposes distinct requirements on reasoning abilities, factual accuracy, and language fluency \cite{chen2024large, survey}. No single model consistently outperforms others in all categories due to differences in optimization strategies and architectural design. A model adept at numerical reasoning may struggle to generate coherent narratives, while a model fine-tuned for conversational fluency may lack precision in coding-related tasks \cite{joshi2024strategic, sinha2024small}.

Moreover, within the same domain, task complexity varies significantly. A business-related query can involve simple data retrieval or complex financial forecasting, each requiring different levels of reasoning and contextual understanding \cite{koncel2023bizbench}. Similarly, coding tasks range from syntax corrections to large-scale algorithm optimization, where different models excel in specific areas \cite{yan2023codescope}.
\begin{figure}[t] 
\centering
\includegraphics[width=1\linewidth]{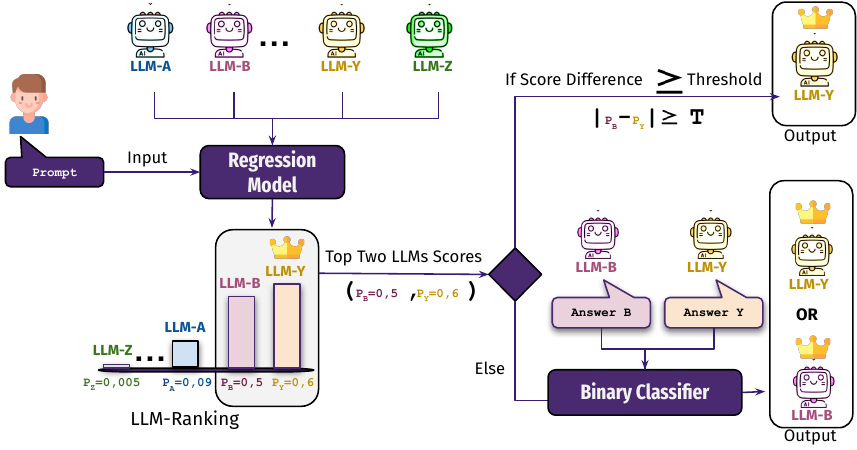} 
\caption{Overview of CARGO. A single embedding regressor scores all candidate LLMs; if the top–two scores are within threshold~\(\tau\), a classifier selects the better one.}
\vspace{-10pt}
 \label{fig:study_method} 
\end{figure}

Most deployments address this challenge by maintaining a portfolio of LLM experts and then routing each query to whichever model promises the best quality \cite{wang2024bench}. However, contemporary routing frameworks exhibit three limitations that undermine their practical utility. 
(1) \textit{Latency overhead}. Some routing frameworks employ sequential cascade strategies: prompts are first processed by small models and, if confidence is insufficient, passed step-by-step to larger ones. This incremental procedure causes latency to accumulate, pushing the worst-case response time toward the sum of all model calls \cite{chen2023frugalgpt}.
(2) \textit{Limited expert models diversity}. Many cost-aware systems reduce routing to a binary choice between one “small” and one “large” model, overlooking the mid-sized, domain-specialist experts that are increasingly common in production environments \cite{ding2024hybrid}.
(3)  \textit{High supervision and scalability constraints } Routers still rely on thousands of labeled prompt–response pairs or reward scores for calibration. Because the cost of label acquisition grows with both the breadth of domains and the size of the expert pool, every expansion or task shift triggers an expensive retraining cycle, hindering operational scalability \cite{wang2024bench, ding2024hybrid}.

Recent benchmarking, further show that existing routing mechanisms, have difficulty generalizing to complex tasks and up-to-date models \cite{hu2024routerbench}.

Motivated by these shortcomings, this paper introduces a confidence-aware routing framework (CARGO) structured around three complementary steps, each carefully designed to resolve key gaps identified in existing approaches. In the first stage, we avoid the common dependence on costly human-annotated data by using candidate LLMs themselves to evaluate response quality. This idea is supported by recent studies demonstrating that LLM-generated assessments closely match human judgements, especially in complex reasoning and knowledge-rich contexts \cite{bavaresco2024llms, verga2024replacing}.
The second stage involves a lightweight embedding-based regression model. This model simultaneously ranks all candidate experts in a single inference pass. Finally, if the two best candidates receive close regression scores, a refinement step invokes a binary classifier trained on pair-wise preferences to break the tie. This overall process is illustrated in Figure~\ref{fig:study_method}.
We provide the source code as a fully reproducible framework to support future research\footnote{\url{https://sites.google.com/view/cascon2025}}.

This work is guided by the following research questions:
\begin{itemize}
    \item \textbf{RQ1 (Individual Model Performance):} How do selected expert LLMs (GPT-4o, Claude 3.5 Sonnet, DeepSeek V3, Perplexity Sonar) individually perform on our curated five-domain dataset?
    \item \textbf{RQ2 (Comparative Routing Efficacy):}  
    How effectively does our confidence-aware routing framework select optimal LLMs compared to individual expert models?
    \item \textbf{RQ3 (Category-Specific Routing Behavior):}  
How does routing performance vary across task categories, and does category-specific modeling further enhance accuracy?

\end{itemize}



\section{Related work}
\label{sec:relatedwork}

The rapid proliferation of large language models (LLMs), accelerated by Transformer architectures, has significantly expanded accessibility but introduced substantial complexity in model selection, with platforms like Hugging Face now hosting over 200,000 models \cite{hari2023tryage}.  As no single model is universally optimal for all tasks \cite{joshi2024strategic}, end-users face considerable difficulty identifying the best fit among diverse architectures, training data, and intended applications. The distinct strengths and weaknesses exhibited by each model necessitate intelligent selection strategies to maximize performance.

Large Language Models (LLMs) often excel in specialized domains when adapted or fine-tuned for specific tasks. For instance, OpenAI’s Codex, fine-tuned for programming, solves 28.8\% of HumanEval coding benchmark problems in a single attempt, whereas the base GPT-3 model solves nearly none \cite{chen2021evaluating}. Similarly, Google’s Minerva, trained on mathematical content, achieves state-of-the-art accuracy on STEM question-answering tasks \cite{dyer2022minerva}, surpassing general LLMs in quantitative reasoning. Instruction-tuned models like InstructGPT \cite{ouyang2022training} further highlight the impact of domain-specific fine-tuning, with a 1.3-billion-parameter variant preferred by human evaluators over the original 175-billion-parameter GPT-3 across diverse prompts.
In other words, no single model performs best across all tasks. Instead, performance depends on training strategies, reasoning capabilities, and task alignment, meaning a model's effectiveness relies on how it was trained, its ability to reason, and how well it fits the requirements of a specific task.

Given the diverse strengths of individual LLMs across tasks, researchers have explored \textit{LLM Ensemble}—a paradigm that strategically integrates multiple models to leverage their complementary capabilities \cite{chen2025ensemble}. Instead of relying on a single model, ensembles aggregate outputs from multiple models to improve accuracy and robustness. 
Basic ensemble methods include majority voting, where outputs from the same or different LLMs are combined, and the final answer is selected by agreement or similarity \cite{li2024more}. Advanced methods such as DEEPEN (Deep Parallel Ensemble) enhance aggregation by integrating token-level predictions from heterogeneous LLMs at each generation step. DEEPEN aligns probability distributions across models by mapping them into a shared representation space based on relative representation theory. The combined predictions are then translated back into the probability space of a designated primary model, allowing the system to handle vocabulary mismatches \cite{huang2024ensemble}.

However, employing ensembles of multiple LLMs can significantly increase computational overhead, latency, and resource demands, especially if models are used indiscriminately. Recent studies, such as Bench-CoE \cite{wang2024bench} and Routing to the Expert \cite{lu2023routing}, explore dynamic routing methods to efficiently manage these demands by selecting the best model or a subset of models based on query characteristics. Bench-CoE \cite{wang2024bench} leverages benchmark evaluations to train routers that assign queries to expert models, while Routing to the Expert \cite{lu2023routing} employs reward-guided mechanisms to route inputs effectively. Despite their dynamic approaches, these methods rely heavily on ground truth datasets to train their routers. This dependency introduces challenges, particularly in acquiring large-scale labeled data, and limits generalization to inputs outside the training data distribution.

Dynamic model selection presents an appealing alternative by intelligently selecting specialized models in real-time. By using lightweight classifiers or meta-models to route queries, dynamic selection efficiently combines accuracy with reduced computational overhead. In this paper, we propose an innovative dynamic routing method that employs LLM-based judges instead of traditional ground truth datasets. We leverage multiple LLM evaluators to train a classifier that dynamically selects the optimal LLM for each input, significantly improving routing effectiveness, particularly in multi-category scenarios.

\section{Methodology}
\label{sec:study-design}

\begin{figure*}[] 
\centering
\includegraphics[width=1\linewidth]{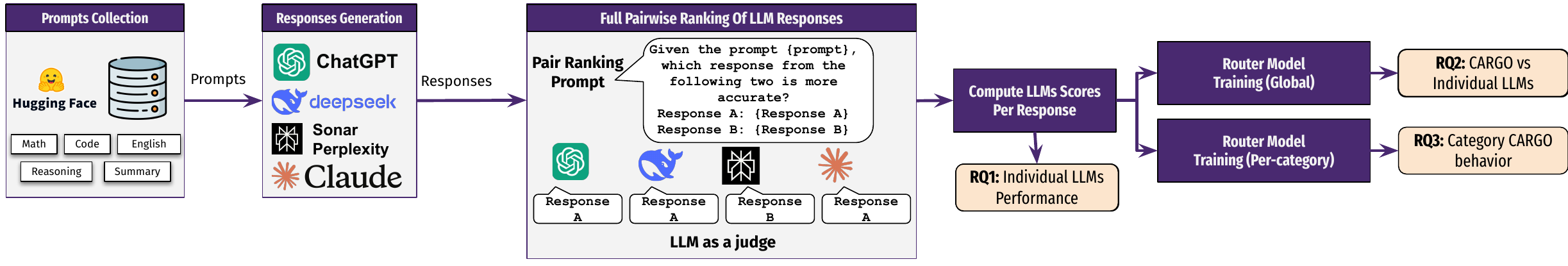} 
\caption{Overview of \textbf{CARGO}. Prompts from five task categories are answered by LLMs. A \emph{pairwise LLM-judging stage} ranks responses and produces labeled data to train confidence-aware regression routers—either \emph{global} or \emph{category-specific}.}

\vspace{-10pt}
\label{fig:study_overview} 
\end{figure*}

We present CARGO, a confidence-aware routing framework that selects the most suitable LLM for each prompt. As shown in Figure \ref{fig:study_overview}: (1) We first collect prompts across diverse domains and obtains responses from multiple LLMs. (2) A pairwise labeling procedure, scores these responses using multiple LLMs as judges. (3) The resulting labeled dataset trains a routing regression models, either global (all domains combined) or category-specific. (4) we finally compares framework performance with individual LLMs.

\subsection{Prompt Dataset Preparation}

Our benchmark draws prompts from Hugging Face datasets in five categories: Mathematics, Coding, Reasoning \& Knowledge, Text Summarization, and Creative Writing.

Each category was initially populated with approximately 1,000 prompts, aggregated from one or more sources. The prompts were shuffled to increase randomness and reduce source-specific patterns. To address overlap across datasets, we applied MinHashLSH \cite{datasketch}, a similarity detection technique, to identify and remove duplicate or highly similar prompts.

After filtering and deduplication, the final prompt counts per category were: 1,085 for mathematics, 1,453 for coding, 958 for reasoning, 1,210 for summarization, and 1,069 for English creative writing. The sets were reshuffled to prevent ordering bias. Table~\ref{tab:datasetssources} summarizes the datasets used to evaluate LLMs across these task categories.

\begin{table}[H]
  \caption{Datasets used for evaluating LLM ranking}
  \label{tab:datasetssources}
  \centering
  \small
  \begin{adjustbox}{max width=\linewidth}
  \begin{tabular}{l l c}
    \toprule
    \textbf{Category} & \textbf{Dataset Source} & \textbf{Subset Size} \\
    \midrule
    Mathematics & 
      \shortstack[l]{OpenThoughts-math \cite{openthoughts_math}, \\ 
                     AI2 ARC \cite{allenai:arc},  
                     MBPP \cite{austin2021program}} 
      & 1,085 \\
    Coding & 
      \shortstack[l]{HumanEval \cite{chen2021evaluating}, \\ 
                     Alpaca \cite{iamtarun_python_code_instructions_18k_alpaca}} 
      & 1,453 \\
    Reasoning \& Knowledge & 
      \shortstack[l]{MMLU-Pro \cite{wang2024mmlu}, \\ 
                     BBH \cite{suzgun2022challenging}} 
      & 958 \\
    Text Summarization & XSum \cite{Narayan2018DontGM} & 1,210 \\
    Creative Writing & 10k Prompts Ranked \cite{10k_prompts_ranked} & 1,069 \\
    \bottomrule
  \end{tabular}
  \end{adjustbox}
\end{table}


\subsubsection{LLM Response Collection}
To obtain responses, each prompt was individually submitted through the OpenRouter API \cite{openrouter_api}, a unified interface that streamlines querying across multiple language models. We configured each query with a \textit{default temperature of ``0.7''}, which strikes a balance between creativity and hallucination \cite{renze2024effect}. We selected four distinct LLMs for their complementary strengths and diverse architectures:
\begin{itemize}
    \item \textbf{ChatGPT-4o}: Chosen for strong general-purpose reasoning and fluent multilingual output~\cite{openai2024gpt4o}.
    
    \item \textbf{Claude 3.5 Sonnet}: Selected for advanced reasoning and large context handling~\cite{anthropic2024claude}.
    
    \item \textbf{DeepSeek V3}: Included for fast inference and factual accuracy via a Mixture-of-Experts (MoE) that activates only a subset of components per query~\cite{deepseek2024v3}.
    
    \item \textbf{Perplexity Sonar}: Used for accurate, concise responses with real-time web retrieval~\cite{perplexity2024sonar}.
\end{itemize}

Each model’s output was captured and stored alongside the corresponding prompt.

\subsection{Pairwise Annotation with LLM Judges}

Inspired by recent ranking methodologies, we propose a pairwise labeling approach, which leverages multiple LLMs as judges. Our approach builds upon traditional pairwise ranking strategies \cite{jiang2023llm, luo2024prp}, incorporating a jury of independent LLM evaluators to reduce bias and improve stability \cite{badshah2024reference, verga2024replacing}.

\subsubsection{Approach Overview}
For each prompt, we collect responses from $M$ distinct LLMs. We then: 
\begin{enumerate}[label=\alph*.]
    \item \textbf{Generate All Pairs:} Construct every possible pair of responses $(R_i, R_j)$. 
    \item \textbf{Multiple LLM Judges:} Present each pair to $N$ different LLM judges, ensuring the judges remain blind to which LLM authored each response. 
    \item \textbf{Comparison Criteria:} Judges compare pairs based on \emph{clarity, accuracy, and completeness}, returning a score indicating which response is preferred. 
    \item \textbf{Score Aggregation:} Pairwise comparisons yield cumulative scores that position each response in a final ranking.
\end{enumerate}

By relying on this llm-based annotation approach rather than on hand-crafted gold answers, the framework can scale naturally to domains where ground truth is ambiguous or absent.

\subsubsection{Scoring Mechanism}
Each judge $J_k$ assigns one of three possible scores $ a_{i,j}^{(k)} $ when comparing response $ R_i $ to $ R_j $:
\begin{equation}
a_{i,j}^{(k)} =
\begin{cases}
1   & \text{if judge $J_k$ prefers response $R_i$,} \\
0.5 & \text{if $R_i$ and $R_j$ are equally good (tie),} \\
0   & \text{if judge $J_k$ prefers response $R_j$.}
\end{cases}
\end{equation}

\noindent For each pair \((R_i, R_j)\), we aggregate judge scores as \(s_{i,j}\):

\begin{equation}
s_{i,j} = \sum_{k=1}^{N} a_{i,j}^{(k)};\quad S_i = \sum_{j \neq i} s_{i,j}.
\end{equation}
Here, \(N\) is the total number of judges. A higher \(s_{i,j}\) indicates that \(R_i\) consistently outperformed \(R_j\) in pairwise evaluations. The global score \(S_i\) for a given response \(R_i\) is derived by summing its pairwise scores against all other responses. A larger \(S_i\) signals stronger overall performance relative to the other \(M-1\) responses.

\subsubsection{Ranking Outcomes}
Finally, we rank all responses in descending order of their total scores $S_i$. The top-scoring response is considered the ``best'' solution for that prompt, followed by others in decreasing order of preference. By incorporating multiple independent judges, we mitigate single-model biases.

\noindent{\textbf{Example:}}
Consider four responses $\{A, B, C, D\}$ to a single prompt. We compare each pair (e.g., $A$ vs.\ $B$, $A$ vs.\ $C$, $A$ vs.\ $D$, etc.), presenting them to four independent LLM judges. If three judges prefer $A$ over $B$ while one finds them equally good, $A$ accumulates 3.5 points and $B$ gets 0.5 for that pair. Repeating this for all six pairs yields a total score per response, which determines the final ranking.

\subsubsection{Evaluation Protocol}\label{sec:evaluation}

We employ three evaluations to assess the reliability and fairness of the pairwise LLM-based annotation scheme using a stratified sample of 250 prompts (50 prompts per category).




\noindent\underline{Inter--judge agreement:}
To assess the consistency of our LLM jury, we compute pairwise Cohen’s~$\kappa$ between all judge pairs and report the mean. Agreement scores range from 59.50\% to 64.06\%, with an average of 62.06\%, corresponding to \(\kappa = 0.62\).
This reflects \textit{substantial} agreement\footnote{Standard interpretation: $\kappa<0.20$ (slight), $0.21$--$0.40$ (fair), $0.41$--$0.60$ (moderate), $0.61$--$0.80$ (substantial), $>0.80$ (near--perfect).}, indicating a stable and reliable ranking signal.

\noindent\underline{Self‐preference bias:}
We assess fairness by measuring the \emph{self-win rate}, the frequency with which a judge selects its own output.  
With four candidates, the expected rate under random selection is \(25\%\).  
Observed rates are: \texttt{gpt-4o} (26.3\%), \texttt{claude-3.5-sonnet} (30.4\%), \texttt{deepseek-chat} (31.1\%), and \texttt{sonar} (24.6\%).  
Overall, results suggest most judges behave fairly, with limited bias.

\noindent\underline{Human validation:}
To assess external validity, two independent professionals reviewed 250 prompts sampled across five categories.  
For each, they selected the best response from four anonymized LLM outputs.  
We compare their choices to the top-ranked outputs from our annotator using Cohen’s~$\kappa$.  
Observed averaged agreement corresponds to \(\kappa = 0.72\).  
These results indicate strong alignment between human judgments and the automatic rankings.


\subsection{Routing Model Training}
We trained regression models to predict the optimal LLM response using two primary strategies: \emph{global} (combining all categories) and \emph{category-specific}. Below, we detail our data preparation method, describe the regression models used, and outline the experimental setup.

\noindent\underline{{Embedding and Scoring Preparation:}}
For each prompt, embeddings were generated using OpenAI's \texttt{text-embedding-ada-002} model. Scores from various LLMs were normalized to sum to 1, enabling fair comparability across prompts and models. These embeddings served as input features for regression models, with the normalized scores used as regression targets.

\noindent\underline{Classification Models:}
We constructed a dataset of LLM comparisons by generating examples where, for each prompt, one model outperformed another. Each entry contains the prompt embedding, metadata for two LLMs, and a label indicating the better one. This dataset was used to train binary classifiers to predict the better of two given LLMs for a specific prompt.

\noindent\underline{Regression Models:}
We employed four regression models to predict LLM scores: Random Forest (RF), Ridge Regression, XGBoost, and a Multi-Layer Perceptron (MLP). All models were trained using Ada embeddings as input features, with normalized LLM scores.

\noindent\underline{Global Model Training:}
In the global training approach, prompts from all categories (mathematics, coding, reasoning, summarization, and English) were combined into a single, unified dataset. Regression models were trained on this combined dataset to predict LLM performance across diverse tasks.

\noindent\underline{Category-Specific Training:}
In the category-specific approach, separate regression models were trained independently for each task category. Each model used only prompts from its respective domain, allowing specialized predictions tailored specifically to that task.

\noindent\underline{Evaluation Methodology:} We partitioned our dataset into 80\% training and 20\% validation sets, employing stratified sampling to preserve category proportions in both sets.

We evaluated the regression models based on their ability to predict LLM performance using the following metrics:
\begin{itemize} 
\item \textbf{Mean Squared Error (MSE):} Quantifies the prediction error between the model's output scores and the actual normalized performance scores for each LLM.

\item \textbf{Accuracy of Predicting the Best Model (Top-1 Accuracy)}: The percentage of prompts for which the regression model correctly identified the top-performing LLM.

\item \textbf{Top-2 Inclusion Rate (Top-1-or-2 Accuracy):} The proportion of prompts for which the true top-performing LLM appeared within the model’s top two predictions.

\item \textbf{Win Rate}: Reflects how frequently the router model either correctly selected the optimal LLM or an LLM with superior actual performance. Formally defined as: 

\begin{equation}
\begin{aligned}
\text{Win Rate} = \frac{1}{\text{Total}} \Big(
    0.5 \times N_\text{spec} 
    + N_\text{better}
\Big)
\end{aligned}
\end{equation}

\noindent
\textbf{Where:}
\begin{itemize}
    \item $N_\text{spec}$: Number of times the specific LLM was picked.
    \item $N_\text{better}$: Number of times top LLM was preferred.
    \item \textit{Total}: Total number of prompts evaluated.
\end{itemize}
\end{itemize}


\noindent\textbf{Example} (CARGO vs. DeepSeek Win Rate): If \textit{DeepSeek} is selected 4 times, better-performing models 3 times, and worse models once out of 8, its win rate: $(4*0.5+3)*100/8=62.5\%$.

\begin{figure*}[ht]
    \centering
    \begin{minipage}{0.32\linewidth}
        \centering
        \includegraphics[width=\linewidth]{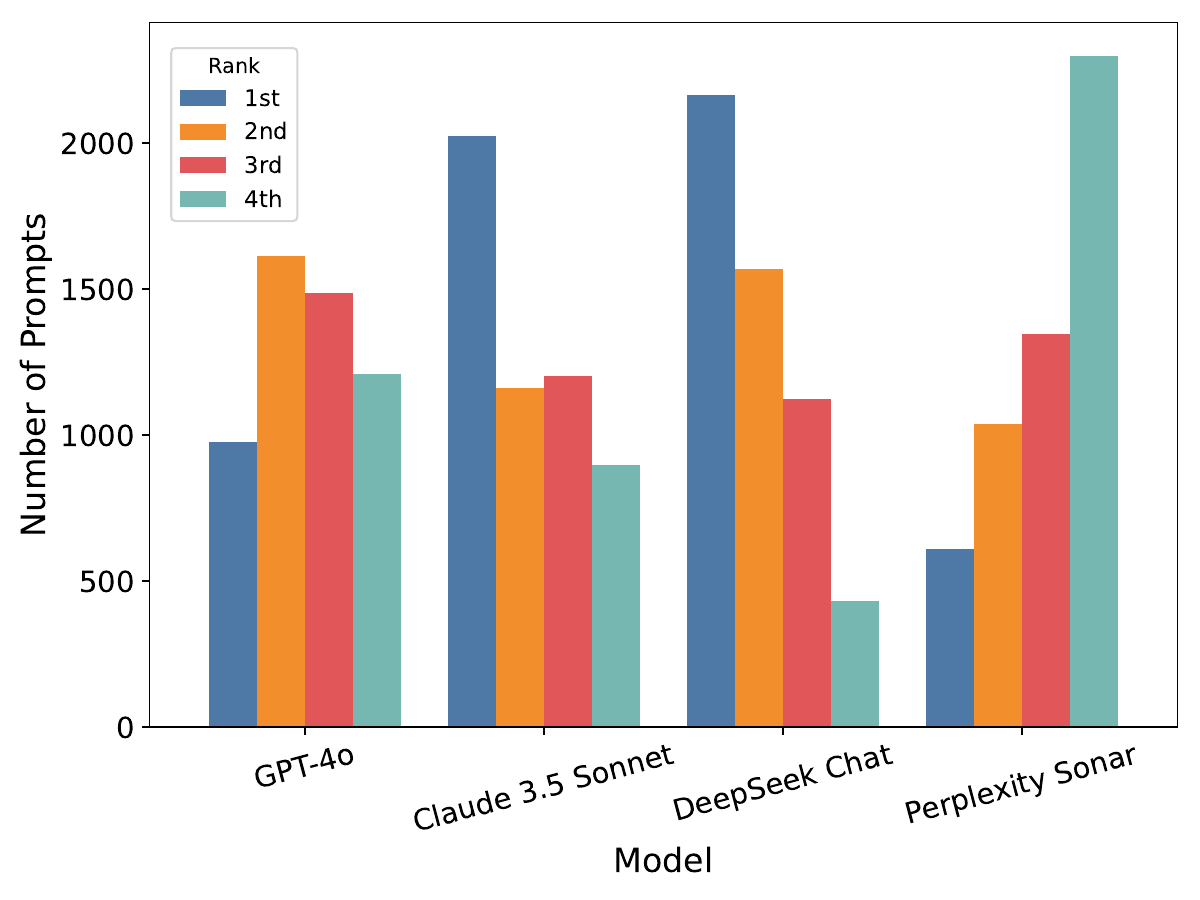}
        {\scriptsize \textbf{(a) All Tasks}}
    \end{minipage}
    \hfill
    \begin{minipage}{0.32\linewidth}
        \centering
        \includegraphics[width=\linewidth]{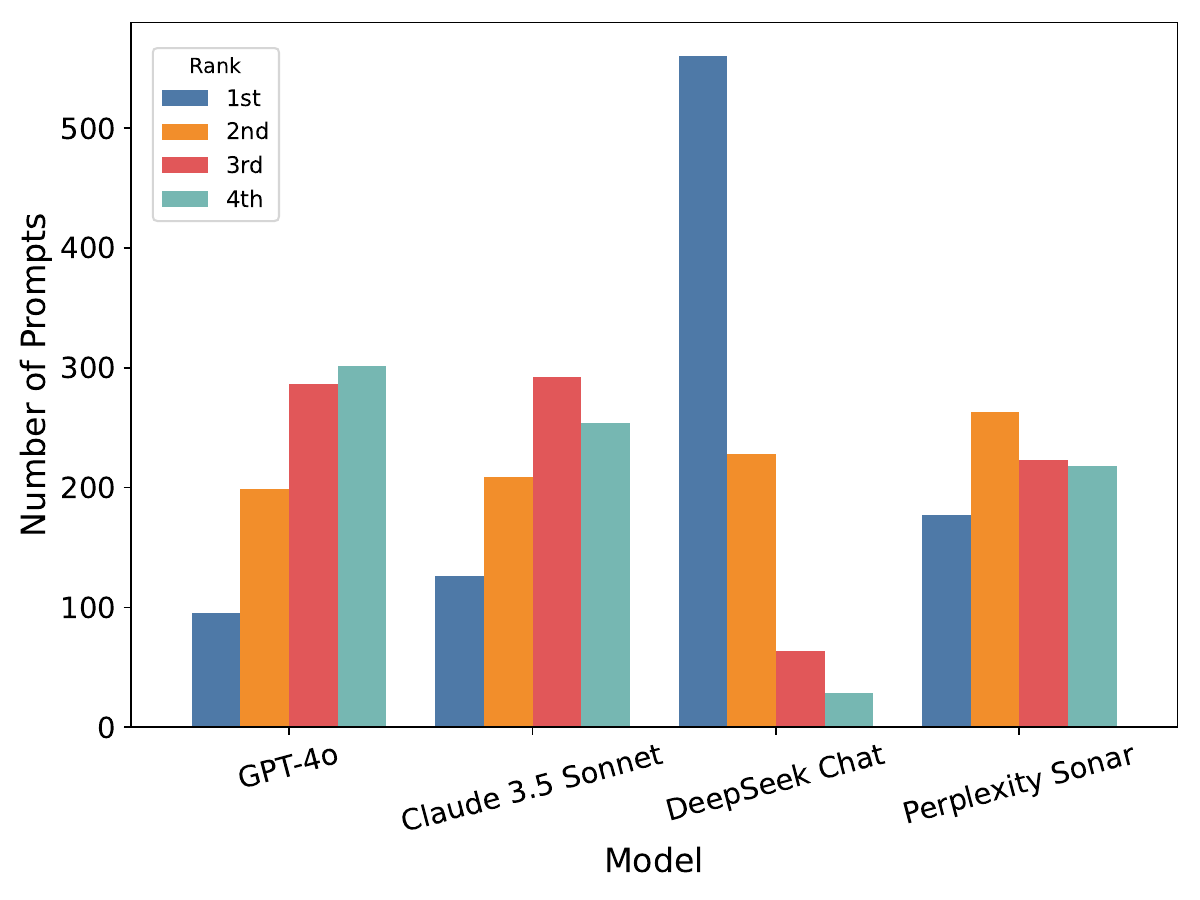}
        {\scriptsize \textbf{(b) Reasoning}}
    \end{minipage}
    \hfill
    \begin{minipage}{0.32\linewidth}
        \centering
        \includegraphics[width=\linewidth]{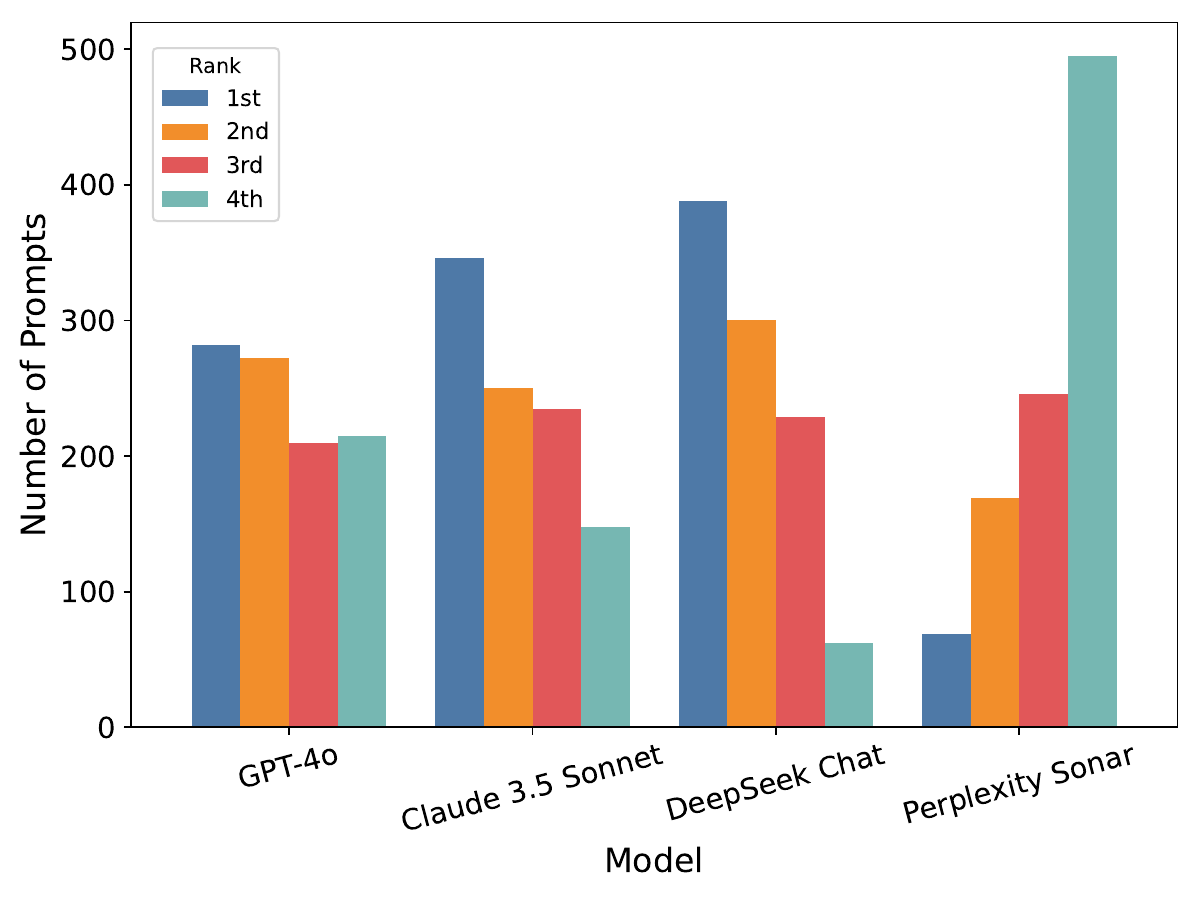}
        {\scriptsize \textbf{(c) Math}}
    \end{minipage}
    
    \vspace{0.3cm}
    
    \begin{minipage}{0.32\linewidth}
        \centering
        \includegraphics[width=\linewidth]{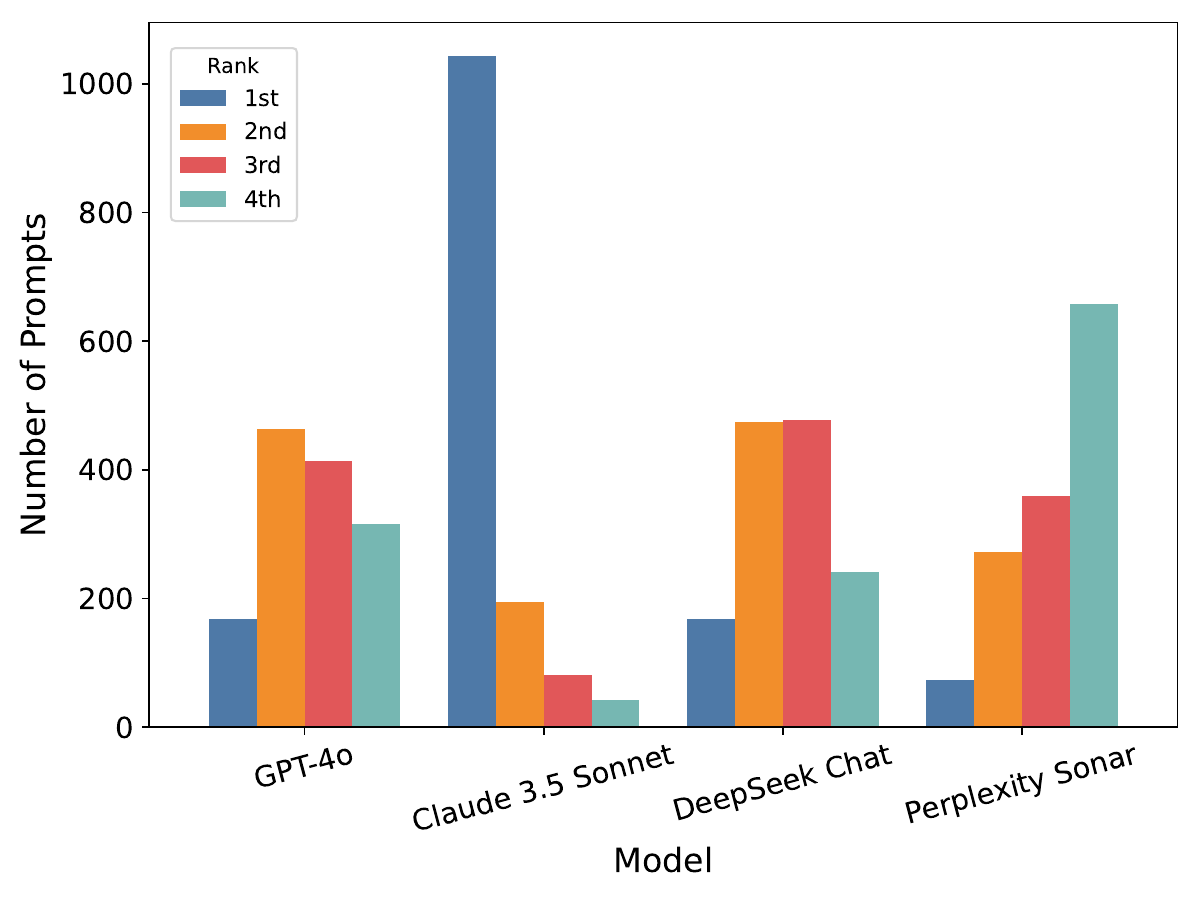}
        {\scriptsize \textbf{(d) Coding}}
    \end{minipage}
    \hfill
    \begin{minipage}{0.32\linewidth}
        \centering
        \includegraphics[width=\linewidth]{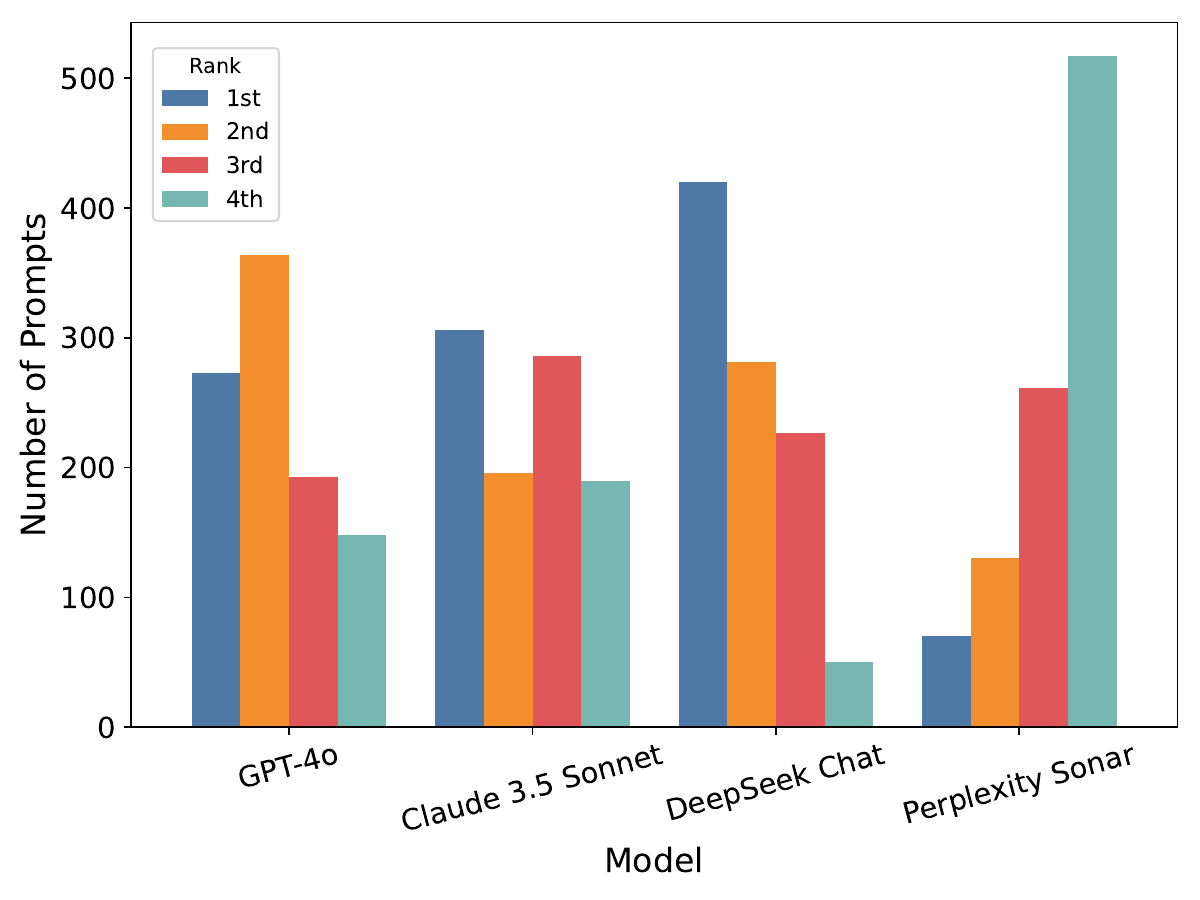}
        {\scriptsize \textbf{(e) English}}
    \end{minipage}
    \hfill
    \begin{minipage}{0.32\linewidth}
        \centering
        \includegraphics[width=\linewidth]{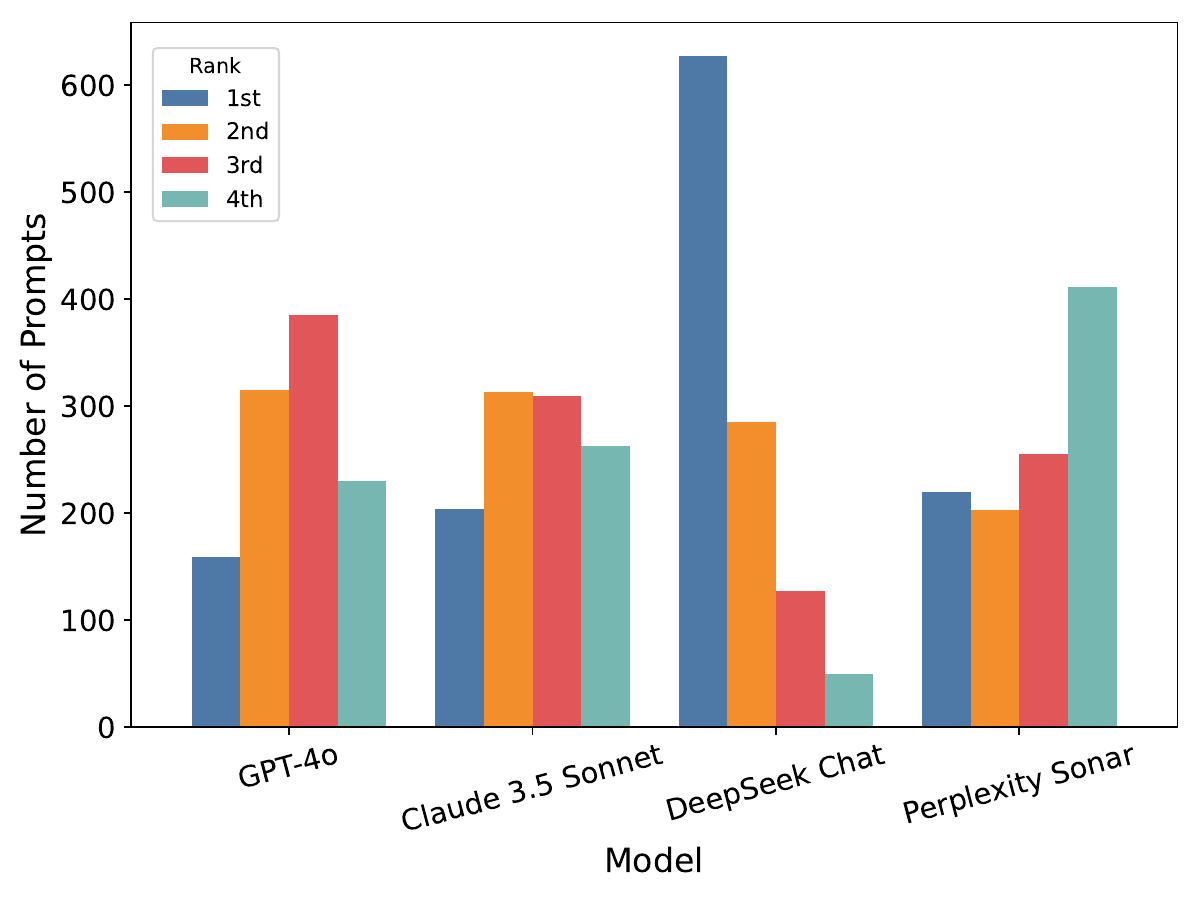}
        {\scriptsize \textbf{(f) Summaries}}
    \end{minipage}
    \caption{Rank frequencies (1 = best, 4 = worst) awarded to each model: (a) all tasks combined, (b) reasoning, (c) math, (d) coding, (e) English-language tasks, and (f) summarization.}
    \label{fig:rank_distributions_combined}
\end{figure*}

\section{Results}
\label{sec:results}

\subsection{RQ1: Task-Specific Performance of Individual LLMs}
This research question examines the task-specific performance and scoring characteristics of four prominent large language models (LLMs): \textbf{GPT-4o}, \textbf{Claude 3.5 Sonnet}, \textbf{DeepSeek}, and \textbf{Sonar}, evaluated over 5,289 prompts across multiple task categories. We present three complementary analyses: (i) distribution of normalized scores by individual models, (ii) frequency of top-ranking performance across different tasks, and (iii) stratification of scores based on assigned ranks.

\begin{figure}[h!]
\centering
\begin{minipage}{0.48\linewidth}
    \centering
    \includegraphics[width=\linewidth]{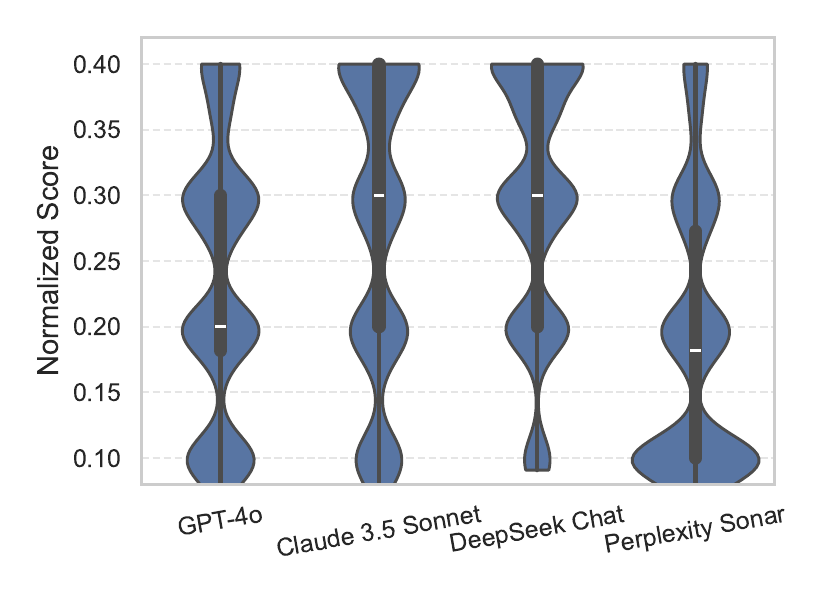}
    \caption*{\small (a) Score distributions per LLM}
\end{minipage}
\hfill
\begin{minipage}{0.48\linewidth}
    \centering
    \includegraphics[width=\linewidth]{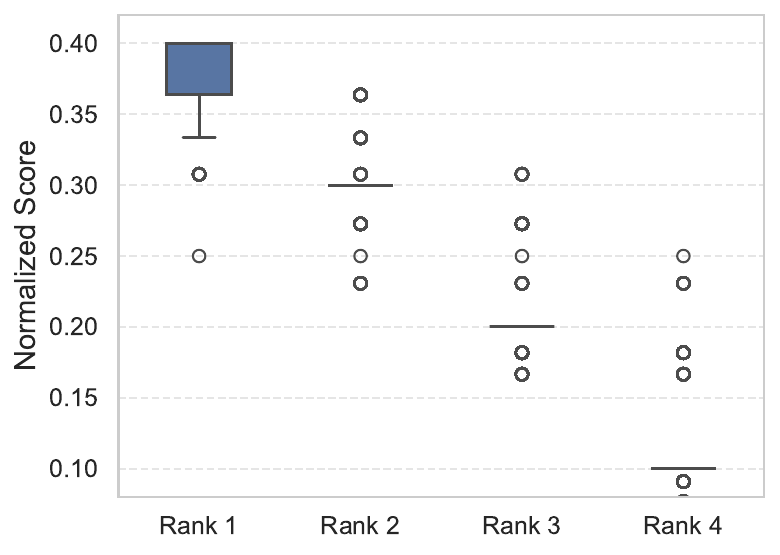}
    \caption*{\small (b) Score distributions by rank}
\end{minipage}
\vspace{-0.5em}
\caption{Normalised Score trends across prompts: (a) by LLM; (b) by rank (Rank-1 to Rank-4).}
\label{fig:normalized_scores}
\vspace{-10pt}
\end{figure}

\noindent\underline{\textbf{Normalized Score Distribution by Model:}}
Figure~\ref{fig:normalized_scores},(a) illustrates the distribution of normalized scores assigned by the PairRanker method for each of the four models across all prompts. The scores for all models fall within a range of 0.08 to 0.40. Distribution patterns emerge, reflecting differences in median performance and score dispersion. Specifically, DeepSeek and Claude 3.5 Sonnet show concentrated high scores, while Sonar exhibits broader, lower distributions.

\noindent\underline{\textbf{Task-Specific Rank Frequencies:}}
We further analyze model performance at a granular task-specific level, reporting how often each model achieved the highest rank. Across the benchmark set, \textbf{DeepSeek} most frequently obtained the top rank (2,164 instances), followed by \textbf{Claude 3.5 Sonnet} (2,025 instances). In contrast, \textbf{GPT-4o} and \textbf{Sonar} were ranked first less often, with 977 and 609 instances, respectively.

In task-specific evaluations, DeepSeek excelled in \textit{reasoning} (560 top ranks) and \textit{math} (388 top ranks). Claude 3.5 Sonnet demonstrated clear dominance in \textit{coding}, achieving the highest rank in 1,043 instances—significantly outperforming other models. In \textit{English-language tasks}, DeepSeek secured the top rank most frequently (420 instances), ahead of Claude 3.5 Sonnet (306), GPT-4o (273), and Sonar, which consistently occupied lower ranks, including 517 instances at rank 4. DeepSeek also emerged as the leader in \textit{summarization}, with 627 top-rank occurrences. Sonar's performance consistently placed it lower across all task categories, particularly in language-intensive evaluations. Detailed rank distributions across task categories are shown in Figure~\ref{fig:rank_distributions_combined}.

\noindent\underline{\textbf{Score Stratification by Rank:}}
Figure~\ref{fig:normalized_scores},(b) analyzes normalized scores across all prompts, stratified by assigned ranks. Rank 1 scores consistently cluster around a median of approximately 0.39. Rank 2 scores exhibit a broader distribution with a median around 0.30, while Ranks 3 and 4 exhibit progressively lower medians. These patterns indicate that models with Rank 1 and Rank 2 typically produce high-quality outputs.

\begin{tcolorbox}[colback=gray!5!white, colframe=black!80!white, title=\textbf{F1: Task-Specific LLM Performance Insights}, fonttitle=\bfseries, boxrule=0.5pt, arc=4pt, left=4pt, right=4pt, top=4pt, bottom=4pt]
\small DeepSeek and Claude outperformed others—\textbf{DeepSeek in reasoning/summarization, Claude in coding}. Scoring shows the top two LLMs achieve a high mean score (0.30–0.39), making them strong candidates per prompt.
\end{tcolorbox}


\subsection{RQ2: Performance of the ensemble vs. individual models}
This research question investigates whether an ensemble-based approach outperforms individual models in predicting the best-performing LLM for a given prompt.

\noindent\underline{\textbf{Classification Models Performance:}}
We evaluated multiple binary classification models to determine their effectiveness in predicting the superior LLM given pairs of models and a prompt embedding. Specifically, we compared Random Forest, Logistic Regression, XGBoost, MLP, and Ridge classifiers, assessing each using accuracy, precision, recall, F1-score, and Area Under the ROC Curve (AUC). Table~\ref{tab:classification_results} summarizes the performance metrics.

The MLP classifier achieved the best performance overall, yielding the highest accuracy (0.826), precision (0.844), recall (0.805), F1-score (0.824), and AUC (0.903). Random Forest and XGBoost followed closely, demonstrating competitive results, while Logistic Regression and Ridge classifiers performed noticeably lower across all metrics.

\begin{table}[htbp]
\centering
\caption{Performance Comparison of Binary Classifiers}
\label{tab:classification_results}
\begin{adjustbox}{width=8.5cm}
\begin{tabular}{lccccc}
\hline
\textbf{Model} & \textbf{Accuracy} & \textbf{Precision} & \textbf{Recall} & \textbf{F1-score} & \textbf{AUC} \\
\hline
Random Forest        & 0.767 & 0.770 & 0.769 & 0.769 & 0.856 \\
Logistic Regression  & 0.643 & 0.650 & 0.638 & 0.644 & 0.685 \\
XGBoost              & 0.763 & 0.774 & 0.752 & 0.763 & 0.846 \\
MLP                  & \textbf{0.826} & \textbf{0.844} & \textbf{0.805} & \textbf{0.824} & \textbf{0.903} \\
Ridge Classifier     & 0.641 & 0.648 & 0.635 & 0.641 & 0.641 \\
\hline
\end{tabular}
\end{adjustbox}
\end{table}

\begin{tcolorbox}[colback=gray!5!white, colframe=black!80!white, title=\textbf{F2: Binary Classification Between Two LLMs}, fonttitle=\bfseries, boxrule=0.5pt, arc=4pt, left=4pt, right=4pt, top=4pt, bottom=4pt]
\small For a given prompt, a binary classifier can determine the better LLM between a pair with up to \textbf{82.6\% accuracy}.
\end{tcolorbox}

\noindent\underline{\textbf{Regression Models Performance:}}
To address this, we trained multiple lightweight regression models, including Random Forest, Ridge Regression, XGBoost, Support Vector Regression (SVR), and Multi-Layer Perceptron (MLP), using a dataset composed of normalized prompt embeddings and corresponding performance scores for GPT-4o, Claude 3.5 Sonnet, DeepSeek, and Perplexity Sonar.

We report each model’s average MSE, Top-1 prediction accuracy, and Top-1-or-2 accuracy, which reflects the proportion of cases in which the model’s top prediction matches either the true best or the true second-best LLM. Table~\ref{tab:regression_results} summarizes the performance of each regression model. Random Forest and Ridge Regression demonstrated the highest Top-1 accuracy, reaching approximately 58\%, while Random Forest also achieved the highest Top-1-or-2 accuracy at 83.65\%.

\begin{table}[htbp]
    \centering
    \caption{Predictive performance of regression models.}
    \label{tab:regression_results}
    \begin{tabular}{lccc}
        \hline
        \textbf{Model} & \textbf{Avg.\ MSE} & \textbf{Top-1 Acc.} & \textbf{Top-1-or-2 Acc.} \\ \hline
        Random Forest  & \textbf{0.0079} & \textbf{58.41\%} & \textbf{83.65\%} \\
        XGBoost        & 0.0082 & 57.84\% & 83.27\% \\
        Ridge          & 0.0079 & 58.13\% & 82.42\% \\
        SVR            & 0.0087 & 56.14\% & 79.77\% \\
        MLP            & 0.0110 & 51.51\% & 75.33\% \\ \hline
    \end{tabular}
\end{table}

\noindent\underline{\textbf{Regression Analysis of Predicted Score Gaps:}}
We analyzed the regression predictions by computing the gap between the top predicted LLM (Rank 1) and the next-best options (Rank 2 and Rank 3). We focused this analysis on the Random Forest model, given its best performance.
We compute the score gaps:
\[
g_{1\!-\!2}(p)=\hat y_{(1)}(p)-\hat y_{(2)}(p),\qquad
g_{1\!-\!3}(p)=\hat y_{(1)}(p)-\hat y_{(3)}(p).
\]

\begin{figure}[htbp]
\centering
\begin{minipage}{0.48\linewidth}
    \centering
    \includegraphics[width=\linewidth]{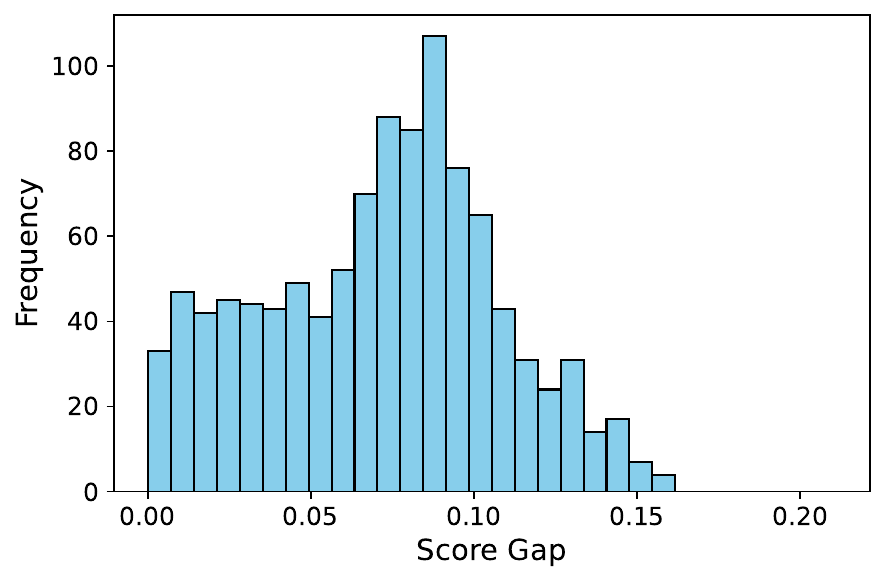}
    \caption*{\small (a) Rank-1 vs Rank-2}
\end{minipage}
\hfill
\begin{minipage}{0.48\linewidth}
    \centering
    \includegraphics[width=\linewidth]{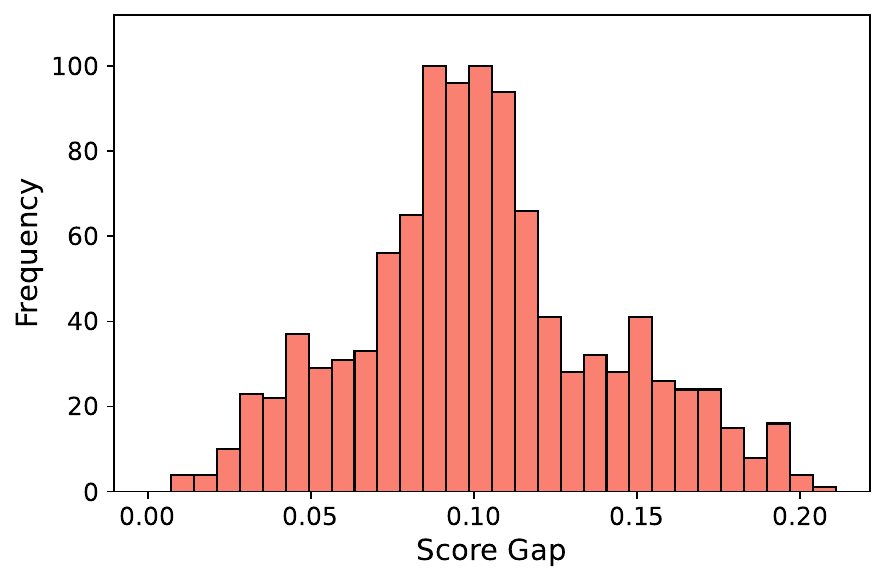}
    \caption*{\small (b) Rank-1 vs Rank-3}
\end{minipage}
\vspace{-0.5em}
\caption{Predicted score gaps from RF regression: (a) top-2 LLMs; (b) top-1 vs. third-ranked LLM.}
\label{fig:score_gap_hists}
\end{figure}
Figure~\ref{fig:score_gap_hists} presents histograms of these predicted score gaps, illustrating how closely the top predictions are spaced.
\textit{The Rank 1 vs Rank 2 gap has a mean of 0.072}, whereas the Rank 1 vs Rank 3 gap averages 0.162. These values show that the regressor assigns its top two candidates close scores.

\begin{tcolorbox}[colback=gray!5!white, colframe=black!80!white, title=\textbf{F3: Regression Performance Analysis}, fonttitle=\bfseries, boxrule=0.5pt, arc=4pt, left=4pt, right=4pt, top=4pt, bottom=4pt]
\small Random Forest achieved the best prediction accuracy among all regressors, with \textbf{top-1 and top-1-or-2 accuracies of 58.41\% and 83.65\%, respectively}. Score gap analysis shows that the \textbf{predicted scores for the top two LLMs are often close}.
\end{tcolorbox}

\noindent\underline{\textbf{Confidence-Based Routing Decision:}}
Our regression analysis revealed that the mean score gap between the top two predicted LLMs is only \(0.07\), suggesting that many prompts yield two closely ranked candidates. To exploit this observation without incurring the cost of always querying multiple models, we introduce a \textit{confidence-based routing policy} guided by the predicted score gap.

For a prompt~\(p\), let \(\hat y_{(1)}(p)\) and \(\hat y_{(2)}(p)\) denote the highest and second-highest scores predicted by the regressor. We define the confidence gap as
\[
g(p) = \hat y_{(1)}(p) - \hat y_{(2)}(p).
\]

Given a threshold~\(\tau\), we apply the following decision rule:

\begin{enumerate}
  \item If \(g(p)\ge\tau\), the router issues a \emph{single} query to the top-ranked LLM.
  \item Otherwise, the router:
  \begin{enumerate}
      \item Sends queries to both the top- and second-ranked LLMs.

      \item Runs a binary classifier to select the superior response from these two queries.
  \end{enumerate}
\end{enumerate}

To identify a suitable \(\tau\), we sweep values from \(0.01\) to \(0.20\) and report the following metrics:
\begin{itemize}
  \item \textbf{Coverage Accuracy}: The percentage of prompts where the set of LLMs chosen for evaluation (usually the top-1 predicted LLM, and the top-2 if the model is uncertain) includes the actual best-performing LLM.
  \item \textbf{Overall Selection Accuracy}: The percentage of prompts where the router's final decision, either the top-1 predicted LLM or the one selected between the top two using a classifier, matches the LLM that actually performed best.

\end{itemize}

We evaluated our confidence-based LLM routing strategy using a two-stage system where a Random Forest regressor first predicts the quality scores of each LLM given a prompt embedding. When the gap between the top two scores falls below a threshold~\(\tau\), a second-stage MLP classifier is invoked to select between them.

\begin{figure}[h!]
    \centering
    \includegraphics[width=0.95\linewidth]{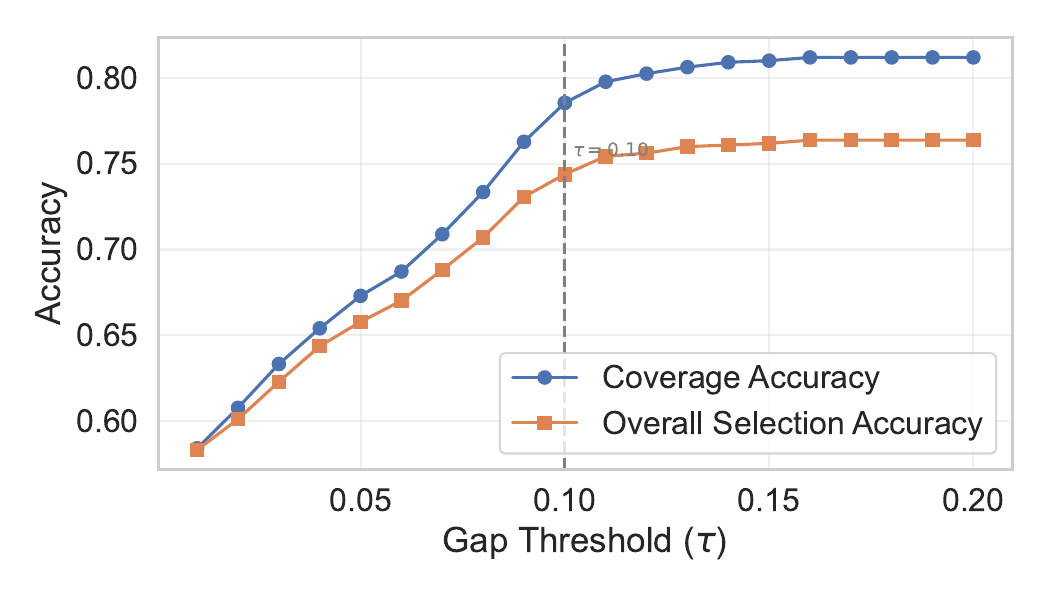}
\caption{Accuracy–Confidence trade-off from variation analysis of the regression gap between the top two predicted LLMs.}

    \label{fig:acc_vs_cost_rf}
\end{figure}

Figure~\ref{fig:acc_vs_cost_rf} shows how performance evolves as \(\tau\) increases. Coverage accuracy improves from 58.4\% at \(\tau = 0.01\) to 78.5\% at \(\tau = 0.10\), while overall selection accuracy rises from 58.3\% to 74.4\% over the same range. Beyond \(\tau = 0.10\), both metrics plateau, with marginal gains up to 81.2\% coverage and 76.4\% accuracy at \(\tau = 0.20\). This trend highlights a favorable trade-off where the regressor is used in confident cases, and the classifier is invoked only in uncertain ones, resulting in strong accuracy without added routing complexity.

\noindent\underline{\textbf{Win Rate Comparison: CARGO vs.\ LLMs:}}
We analyse the win rate, which quantifies how often a model selects an LLM that performs better or equally well compared to a reference LLM on a given prompt. It reflects the model’s ability to identify the most suitable LLM response.
\textit{Ensemble models consistently achieve higher or comparable win rates than any single LLM across all evaluated cases.} As shown in Table~\ref{table:win_rates}, Random Forest achieves the highest win rates among regressors for \texttt{gpt-4o} (72.87\%) and \texttt{perplexity} (84.12\%), and ties for the best result on \texttt{deepseek} (61.15\%). Once combined with the MLP classifier, the pipeline further improves performance. At \(\tau=0.12\), it reaches 82.28\%, 73.53\%, 71.36\%, and 89.18\% respectively across the four LLMs, with an average win rate of 79.09\%. These results confirm that the two-stage pipeline is more effective at selecting the better LLM than any standalone model or one-stage ensemble.


\begin{table}[h]
\centering
\caption{Win rates (\%) of CARGO models versus each LLM.}
\label{table:win_rates}
\begin{tabular}{lcccc}
\toprule
\textbf{Model} & \textbf{GPT-4o} & \textbf{Claude 3.5} & \textbf{DeepSeek} & \textbf{Perplexity} \\
\midrule
\multicolumn{5}{c}{\emph{Regressor‑Only Models}} \\
Random Forest & 72.87 & 63.42 & 61.15 & 84.12 \\
XGBoost       & 73.11 & 63.33 & 60.59 & 83.41 \\
Ridge         & 73.25 & 63.33 & 60.26 & 83.13 \\
SVR           & 71.25 & 61.81 & 59.22 & 81.47 \\
MLP           & 69.42 & 59.83 & 55.39 & 78.12 \\
\midrule
\multicolumn{5}{c}{\emph{CARGO Routing Pipelines (RF + MLP)}} \\
$\tau = 0.07$ & 78.69 & 70.04 & 67.67 & 87.19 \\
$\tau = 0.10$ & 81.62 & 72.92 & 70.79 & 88.94 \\
$\tau = 0.12$ & \textbf{82.28} & \textbf{73.53} & \textbf{71.36} & \textbf{89.18} \\
\bottomrule
\end{tabular}
\end{table}


 \begin{tcolorbox}[colback=gray!5!white, colframe=black!80!white, title=\textbf{F4: Confidence-Based Routing Performance}]
\small The confidence-based two-stage pipeline (Random Forest regressor + MLP classifier) substantially improves LLM selection, achieving \textbf{76.4\% overall accuracy} at $\tau=0.10$. It also consistently outperforms any single LLM, with a \textbf{mean win rate of 79.1\%} across the expert pool.
\end{tcolorbox}

\subsection{RQ3: Category-Specific Ensemble Task Behavior}
This section investigates how ensemble-based routing strategies perform across different task categories. By analyzing both regression and classification results, we aim to understand whether certain categories benefit more from ensemble decision

\noindent\underline{\textbf{Category Classification Accuracy:}}
We trained a Random Forest to predict each prompt’s high-level category (Coding, English, Math, Reasoning, Summaries). The classifier achieved 96 \% accuracy on held-out data, with overall precision, recall, and F1 all approximately 0.96 (see Table \ref{tab:category_results}).

\begin{table}[ht]
\centering
\caption{Random Forest category classification performance}
\label{tab:category_results}
\begin{tabular}{lcccc}
\toprule
\textbf{Category} & \textbf{Precision} & \textbf{Recall} & \textbf{F1-score}\\
\midrule
Coding     & 0.97 & 0.99 & 0.98  \\
English    & 0.88 & 0.92 & 0.90  \\
Math       & 0.99 & 0.95 & 0.97 \\
Reasoning  & 0.98 & 0.93 & 0.96  \\
Summaries  & 0.98 & 0.99 & 0.98  \\
\midrule
\textbf{Overall} & \multicolumn{3}{c}{\textbf{Accuracy = 0.9603}}  \\
\bottomrule
\end{tabular}
\end{table}

\noindent\underline{\textbf{Regression Model Performance by Category:}}
As shown in Table~\ref{tab:regression_performance}, regression models exhibit consistent yet varied performance across the five task categories. Among them, \textbf{Ridge Regression} and \textbf{Random Forest} emerge as the most stable performers, both achieving strong average MSE.

\textbf{Top-1 Accuracy} varies more significantly across categories. While models may not always precisely identify the single best-performing LLM, they remain consistently effective at narrowing down the top two candidates. This is reflected in the noticeably higher Top-1-or-2 Accuracy across all categories. Notably, tasks such as \emph{Coding} and \emph{Reasoning} exhibit both high Top-1-or-2 Accuracy (up to 0.919 and 0.864, respectively) and relatively strong Top-1 Accuracy, with values exceeding 0.74 and 0.63 in the best cases.

\begin{table}[h]
\centering
\caption{Regression Model Performance by Category}
\label{tab:regression_performance}
\begin{adjustbox}{width=8.5cm}
\begin{tabular}{lcccc}
\toprule
\textbf{Category} & \textbf{Model} & \textbf{Avg. MSE} & \textbf{Top-1 Acc.} & \textbf{Top-1-or-2 Acc.} \\
\midrule
\multirow{5}{*}{Summaries}
    & Random Forest   & 0.0093 & \textbf{0.518} & \textbf{0.748} \\
    & XGBoost         & 0.0095 & 0.509 & 0.743 \\
    & MLP             & 0.0120 & 0.413 & 0.642 \\
    & \textbf{Ridge}  & \textbf{0.0092} & 0.514 & \textbf{0.748} \\
    & SVR             & 0.0098 & 0.472 & 0.706 \\
\midrule
\multirow{5}{*}{English}
    & Random Forest   & 0.0078 & 0.500 & 0.750 \\
    & XGBoost         & 0.0081 & 0.485 & 0.719 \\
    & MLP             & 0.0099 & 0.459 & 0.704 \\
    & \textbf{Ridge}  & \textbf{0.0076} & 0.500 & \textbf{0.770} \\
    & SVR             & 0.0085 & 0.459 & 0.745 \\
\midrule
\multirow{5}{*}{Math}
    & Random Forest   & 0.0079 & 0.474 & 0.745 \\
    & XGBoost         & 0.0083 & 0.454 & \textbf{0.750} \\
    & \textbf{MLP}    & \textbf{0.0077} & \textbf{0.510} & \textbf{0.750} \\
    & \textbf{Ridge}  & \textbf{0.0077} & 0.490 & 0.735 \\
    & SVR             & 0.0089 & 0.495 & 0.765 \\
\midrule
\multirow{5}{*}{Coding}
    & \textbf{Random Forest}   & \textbf{0.0062} & 0.747 & 0.912 \\
    & XGBoost         & 0.0066 & 0.747 & 0.912 \\
    & MLP             & 0.0064 & \textbf{0.762} & \textbf{0.919} \\
    & \textbf{Ridge}  & \textbf{0.0062} & 0.755 & 0.905 \\
    & SVR             & 0.0074 & 0.740 & \textbf{0.919} \\
\midrule
\multirow{5}{*}{Reasoning}
    & Random Forest   & 0.0073 & 0.616 & 0.859 \\
    & XGBoost         & 0.0077 & \textbf{0.638} & 0.859 \\
    & MLP             & 0.0078 & 0.621 & 0.853 \\
    & Ridge           & \textbf{0.0076} & 0.627 & 0.859 \\
    & SVR             & 0.0084 & \textbf{0.638} & \textbf{0.864} \\
\bottomrule
\end{tabular}
\end{adjustbox}
\end{table}

\noindent\underline{\textbf{Confidence-Based Routing Decision by Category:}}
We aim to improve the prediction accuracy of the top candidate LLM selected to answer a given prompt per category. Specifically, we analyze how the routing behavior varies across different macro-categories based on the predicted score gap between the top two LLM candidates. We use the Random Forest regressor as the primary router, since Random Forest performed consistently well across all categories in earlier evaluations. The MLP classifier applied in this routing pipeline is the same as in RQ2, trained once globally on the full dataset.

\begin{figure}[]
    \centering
    \includegraphics[width=0.85\linewidth]{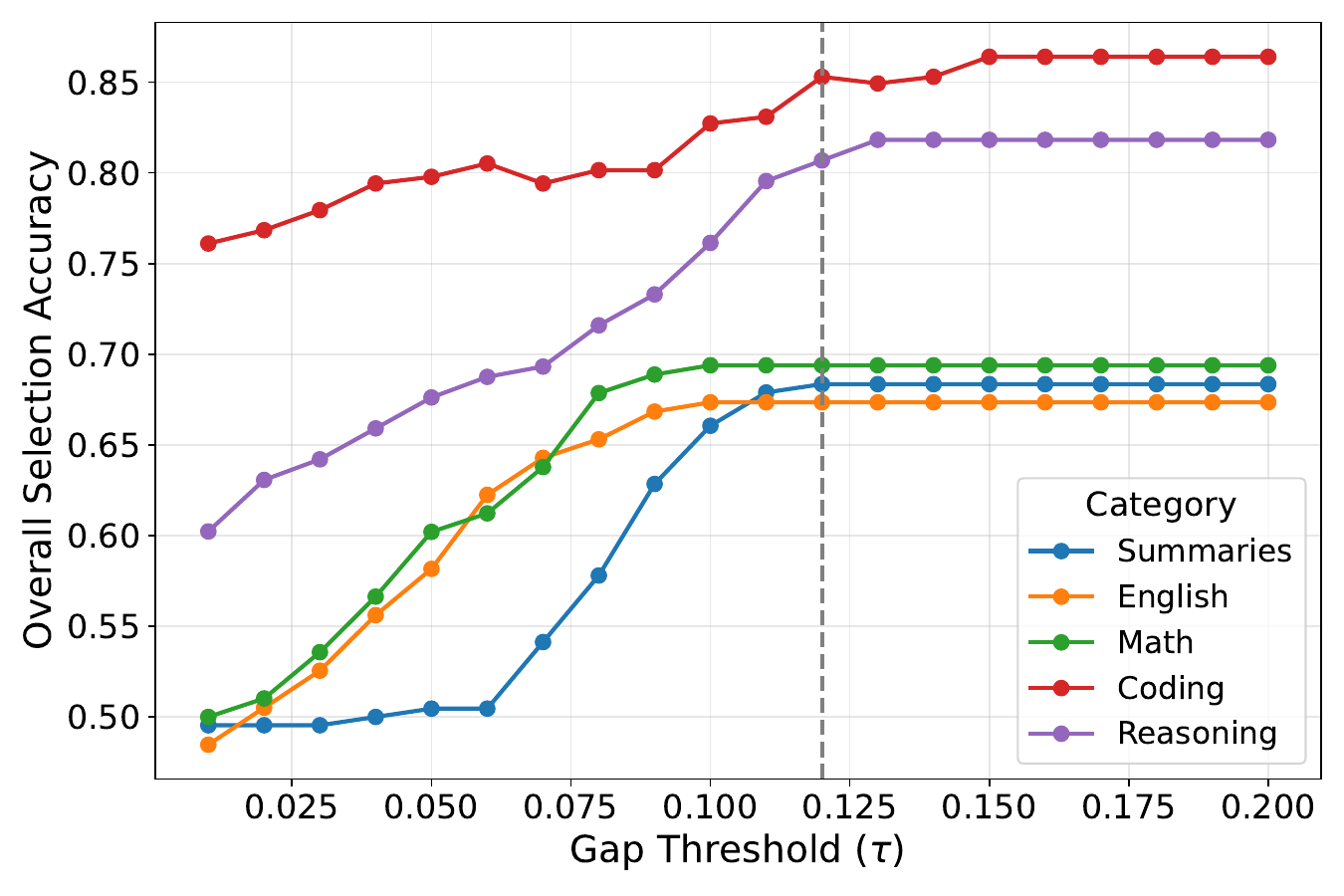}
    \caption{Overall Selection Accuracy vs.\ score gap threshold~\(\tau\) for each prompt category.}
    \label{fig:category_routing_accuracy}
\end{figure}

As shown in Figure~\ref{fig:category_routing_accuracy}, increasing the threshold~\(\tau\) steadily improves the Overall Selection Accuracy in all categories up to a saturation point. Most categories exhibit a rapid gain in performance between \(\tau = 0.05\) and \(\tau = 0.12\), after which the improvement flattens. The \emph{Coding} and \emph{Reasoning} categories achieve the highest accuracy, peaking at 0.864 and 0.818 respectively. In contrast, \emph{Summaries} and \emph{English} show more gradual increases and lower plateaus, reaching maximum accuracies of 0.683 and 0.673 respectively.

\begin{tcolorbox}[colback=gray!5!white, colframe=black!80!white, title=\textbf{F5: Ensemble Pipeline Performance by Category}]
\small 
It was straightforward to predict the category of a given prompt, achieving \textbf{96\% accuracy} using a Random Forest classifier. Although the regression models alone were not always sufficient to reliably select the best-performing LLM, \textbf{Top-1-or-2 Accuracy reached 86.4\% for Coding and 81.8\% for Reasoning}. \textbf{Overall Selection Accuracy increased consistently with a higher score gap threshold~\(\tau\), reaching \(86.4\%\) in coding and \(81.8\%\) in reasoning}.
\end{tcolorbox}

\section{Discussions}
\label{sec:discussions}






\subsection{Accuracy \emph{vs.} Computational Cost}
\label{sec:acc_cost}

To clarify when lightweight routing models (regressors and binary classifiers) should be invoked, we relate their execution cost to the resulting selection accuracy. In our setup, each prompt always goes through a single regressor $R$, which scores all candidate LLMs. A binary classifier $B$ is optionally used to resolve close decisions between the top two candidates. 

Let $g(p)$ denote the score difference between the top-1 and top-2 LLMs predicted by $R$ on prompt $p$. The classifier is called only if $g(p) < \tau$, where $\tau$ is a user-defined threshold controlling the router's conservativeness. The expected number of lightweight model calls per prompt is then:

\begin{equation}
\mathbb{E}[C(\tau)] \;=\; 
C_{R}\;+\;\underbrace{\Pr\!\bigl[g(p)<\tau\bigr]}_{\text{classifier usage}}\cdot C_{B},
\label{eq:cost}
\end{equation}

where $C_R = 1$ and $C_B = 1$ are unit costs for the regressor and the classifier. The classifier usage term is measured empirically on the test set.

\vspace{3pt}
\noindent\underline{\textbf{Global results.}} For the global regressor, increasing $\tau$ causes the router to fall back to the binary classifier more often, leading to higher accuracy. Between $\tau = 0.04$ and $\tau = 0.08$, the classifier is invoked on 23\% to 55\% of prompts, and accuracy rises from 64.4\% to 70.7\%. This is the most efficient part of the curve, where each additional classifier call yields a meaningful gain. After \(\tau = 0.10\), accuracy gains slow down while cost rises sharply. At \(\tau = 0.13\), accuracy reaches 75.9\% with over 80\% classifier usage, and saturates near 76\% beyond that. Using the classifier on half the prompts captures most of the benefit, making \(\tau \in [0.06, 0.10]\) a practical trade-off.

\vspace{2pt}
\noindent\underline{\textbf{Per-category results.}} Using a separate regressor per category improves the balance between accuracy and classifier usage. At \(\tau = 0.13\), \textit{Coding} reaches 84.9\% accuracy with 79\% classifier usage, and \textit{Reasoning} reaches 81.8\% with 94\%. In contrast, \textit{Summaries} and \textit{English} need the classifier on nearly every prompt but still stay below 70\% accuracy. This highlights differences in how well the regressor can distinguish top candidates across domains.

\vspace{2pt}

\noindent\underline{\textbf{Baselines.}} We set baseline models for comparison. The \emph{\textbf{Top-1}} method simply selects the highest-scoring LLM from $R$; it is cheap but often inaccurate. The \emph{\textbf{Top-1 $\vee$ Top-2}} strategy measures how often the true best LLM appears among the top two predictions. While this shows that the regressor can often narrow down the correct choice to a small set, it does not resolve which of the two to pick—choosing at random still risks failure in half the cases. A classifier is therefore essential to make accurate selections within this narrowed set.  
\emph{\textbf{All-pairs voting}} serves as a strong baseline, invoking the classifier \(B\) on every pair of candidate LLMs and selecting the model with the most wins. For \(M = 4\), this results in \(\binom{4}{2} = 6\) classifier calls per prompt. While this approach achieves competitive accuracy, its cost scales quadratically with \(M\), making it impractical for large ensembles and potentially introducing delays that violate service-level agreements (SLAs).

Table~\ref{tab:acc_cost_table} compares all strategies in terms of selection accuracy and expected lightweight cost per prompt, assuming \(M = 4\) candidate LLMs. Both global and category-aware versions of CARGO offer a balanced trade-off, capturing most of the gains of Top-2 or all-pairs voting while requiring far fewer classifier calls.

\begin{table}[h]
\centering
\caption{Accuracy and expected number of lightweight model calls per prompt (4 LLMs). 
}
\label{tab:acc_cost_table}
\begin{tabular}{lcc}
\toprule
\textbf{Routing Strategy} & \textbf{Expected Calls} & \textbf{Accuracy} \\
\midrule
Top-1 only (global $R$)               & $1R$                  & 58.4\,\% \\
Top-1 or Top-2 (accept both)          & $1R$                  & 83.7\,\% \\
All-pairs voting ($\binom{4}{2} = 6B$) & $6B$                 & 82.6\,\% \\
\addlinespace
Global CARGO ($\tau = 0.13$)          & $1R + 0.946B$         & 75.99\,\% \\
\addlinespace
\textit{Per-category CARGO ($\tau = 0.13$):} & & \\
\quad Summaries                        & $1R + 1.000B$         & 68.3\,\% \\
\quad English                          & $1R + 0.995B$         & 67.3\,\% \\
\quad Math                             & $1R + 1.000B$         & 69.4\,\% \\
\quad Coding                           & $1R + 0.794B$         & \textbf{84.9\,\%} \\
\quad Reasoning                        & $1R + 0.938B$         & \textbf{81.8\,\%} \\
\bottomrule
\end{tabular}
\end{table}

\subsection{Scalability: New Experts and New Domains}\label{sec:scalability}

A practical router should scale with expanding model portfolios and evolving tasks with minimal retraining. To quantify this scalability, we conduct two ablation studies that extend the original benchmark: \textbf{(i)}~adding a fifth expert and \textbf{(ii)}~introducing an unseen domain.


\noindent\underline{\textbf{Experiment 1: Adding a Fifth Expert.}}  
We added \textit{Gemini-2.5-Pro} to the original set of four expert models, generated responses for 250 prompts, and re-labeled the resulting pairs. Adding a fifth model introduces 4 new pairs per prompt, each judged by 4 LLMs, resulting in \(250 (prompts) \times 4 (pairs) \times 4 (judges)  = 4{,}000\) additional comparisons. The average Cohen’s kappa remained at \(0.60\), and human agreement with the new top-ranked model held steady at \(71\%\).



\noindent\underline{\textbf{Experiment 2: Adding a sixth domain.}}
We added a \textit{Translation} slice with 50 prompts from SoftAge-AI\footnote{\url{https://huggingface.co/datasets/SoftAge-AI/prompt-eng_dataset}}.  
All four experts produced responses, and each prompt generated 6 unique pairs, judged by 4 LLMs, yielding \(50 \times 6~\text{(pairs)} \times 4~\text{(judges)} = 1{,}200\) additional comparisons.  
Judge agreement for translation was \(\kappa = 0.77\), and human alignment with LLM ranking was \(\kappa = 0.67\), comparable to earlier domains.

These results show that \textit{CARGO} accommodates both portfolio and task expansion without extra human labeling, while preserving agreement, fairness, and alignment. Such plug-and-play scalability is essential for real-world deployments with evolving model inventories and workloads.

\section{Threats To Validity}\label{sec:validity}
\noindent\textbf{Internal Validity.} Our ranking methodology revealed that some models consistently outperformed others, leading to imbalanced rank distributions. This poses a threat to internal validity by potentially biasing the learned models toward dominant LLMs. To reduce this effect, we employed stratified sampling to preserve category and rank diversity in training and test sets.

\noindent\textbf{Construct Validity.} Relying on a single LLM to evaluate responses can introduce self-preference bias, as shown in \cite{wang2023large}. To mitigate this, we used multiple independent LLMs as judges for pairwise comparisons, with final decisions made by majority vote to prevent any model from evaluating its own output. We also reported inter-judge agreement to assess consistency, quantified self-preference bias across models, and validated our ranking methodology through human agreement studies, thereby strengthening the reliability of the evaluation.

\noindent\textbf{External Validity.} While our analysis is based on a curated dataset, there remains a risk that the findings may not generalize to other domains or prompt distributions. To mitigate this, we collected prompts from multiple sources and ensured coverage across diverse categories, increasing the likelihood that our results extend beyond a single task or domain. Additionally, the list of expert LLMs was selected in early 2025 based on then-current market leaders. As the LLM landscape evolves rapidly, newer models may outperform our selected experts. However, our methodology is model-agnostic and remains applicable regardless of which LLMs are included in the expert pool.

\noindent\textbf{Conclusion to Validity.} Since our evaluation relies on LLM-based judges, ensuring generalizability is essential. To address this, we tested the scalability of our system by adding a sixth domain and new expert models. The router maintained consistent performance, indicating robustness to domain and model change.

\section{Conclusion}\label{sec:conclusion}


We introduced \textbf{CARGO}, a lightweight, confidence-aware routing framework for dynamically selecting the most suitable large language model (LLM) per prompt. CARGO integrates an embedding-based regressor with a fallback binary classifier, enabling cost-sensitive control over routing while supporting both global and category-specific configurations.
CARGO was evaluated on four state-of-the-art expert LLMs across five task domains. It achieved a top-1 routing accuracy of 76.4\% globally and up to 86\% in specific domains. Against expert models, our two-stage pipeline consistently outperformed individual LLMs, achieving a mean win rate of 79.1\%.

Unlike prior routing systems that depend on large transformer-based routers (e.g., BERT) and extensive labeled datasets, CARGO relies on lightweight models and LLM-based judgments, requiring significantly less training data. We further demonstrated the robustness of CARGO by extending it to a new domain and adding an expert model, while maintaining performance and agreement levels. These results show that accurate and adaptive LLM selection can be achieved with minimal overhead, making CARGO well-suited for real-world, performance-sensitive deployments where routing speed, model cost, and adaptive accuracy are essential.

\section{Acknowledgement}\label{sec:ack}
This work was conducted in collaboration with \textbf{zuvu.ai}, an industrial partner developing AI-powered tools.




\balance
\bibliographystyle{IEEEtran}
\bibliography{references} 
\end{document}